\documentclass[10pt,letterpaper]{article}
\usepackage[top=0.85in,left=2.75in,footskip=0.75in]{geometry}

\usepackage{amsmath,amssymb}

\usepackage{changepage}

\usepackage{textcomp,marvosym}

\usepackage{cite}

\usepackage{nameref,hyperref}

\usepackage{microtype}
\DisableLigatures[f]{encoding = *, family = * }

\usepackage[table]{xcolor}

\usepackage{array}
\usepackage{multirow}

\newcolumntype{+}{!{\vrule width 2pt}}

\newlength\savedwidth

\newcommand\thickhline{\noalign{\global\savedwidth\arrayrulewidth\global\arrayrulewidth 2pt}%
\hline
\noalign{\global\arrayrulewidth\savedwidth}}

\raggedright
\usepackage[aboveskip=1pt,labelfont=bf,labelsep=period,justification=raggedright,singlelinecheck=off]{caption}

\makeatletter
\renewcommand{\@biblabel}[1]{\quad#1.}
\makeatother

\usepackage{lastpage,fancyhdr,graphicx}
\usepackage{epstopdf}
\fancyheadoffset[L]{2.25in}
\fancyfootoffset[L]{2.25in}
\newcommand{\bn}{\mathbf{n}}

\newcommand{\beq}{\begin{equation}}
\newcommand{\eeq}{\end{equation}}

\begin{document}
\vspace*{0.2in}

\begin{flushleft}
{\Large
\textbf\newline{Machine learning topological defects in confluent tissues} 
}
\newline
\\
Andrew Killeen\textsuperscript{1*},
Thibault Bertrand\textsuperscript{2\dag},
Chiu Fan Lee\textsuperscript{1\ddag},
\\
\bigskip
\textbf{1} Department of Bioengineering, Imperial College London, South Kensington Campus, London SW7 2AZ, U.K.
\\
\textbf{2} Department of Mathematics, Imperial College London, South Kensington Campus, London SW7 2AZ, U.K.
\\
\bigskip

*\, a.killeen18@imperial.ac.uk

\dag\, t.bertrand@imperial.ac.uk

\ddag\, c.lee@imperial.ac.uk

\end{flushleft}

\section*{Abstract}
Active nematics is an emerging paradigm for characterising biological systems. One aspect of particularly intense focus is the role active nematic defects play in these systems, as they have been found to mediate a growing number of biological processes. Accurately detecting and classifying these defects in biological systems is, therefore, of vital importance to improving our understanding of such processes. While robust methods for defect detection exist for systems of elongated constituents, other systems, such as epithelial layers, are not well suited to such methods. Here, we address this problem by developing a convolutional neural network to detect and classify nematic defects in confluent cell layers. Crucially, our method is readily implementable on experimental images of cell layers and is specifically designed to be suitable for cells that are not rod-shaped. We demonstrate that our machine learning model outperforms current defect detection techniques and that this manifests itself in our method requiring less data to accurately capture defect properties. This could drastically improve the accuracy of experimental data interpretation whilst also reducing costs, advancing the study of nematic defects in biological systems. 

\section*{Author summary}
Defects in the local alignment of cells have been found to play a functional role in homeostatic and morphogenetic processes in many different biological systems. Detecting these defects is, therefore, very important for improving our understanding of these processes. However, current defect detection techniques are not well suited to cell layers in which cells are not elongated in shape, and so the direction of cell orientation can be poorly defined, even though defects mediate important homeostatic processes in these layers. Here, we address this problem and develop a machine learning method to detect and classify defects which is specifically designed for systems for which existing methods are not appropriate. We show that our method outperforms current techniques, detecting defects in confluent cell layers more accurately. We then demonstrate that this improved performance means that properties of these defects, often the target of experimental studies, can be characterised more accurately with less data. We anticipate this could drastically improve experiments investigating defects in these systems, improving our knowledge of important biological processes.


\section*{Introduction}
 Tissue dynamics underpins a wide variety of biological processes such as wound healing \cite{Poujade2007}, cancer metastasis \cite{Friedl2009} and  morphogenesis \cite{Weijer2009}. Many of these processes concern confluent tissues, such as epithelial and endothelial cell layers, making suitable descriptions of the dynamics of these systems a prerequisite for our understanding of these processes. Unlike constituents in a passive material, cells within a confluent tissue can generate forces and exert stresses on their neighbours and underlying substrate. As such, active matter physics provides a natural framework for describing confluent tissues and has provided numerous insights into these systems \cite{Alert2020}. Active matter is an emergent field of physics concerned with describing many-body systems far from equilibrium, where the system is driven from equilibrium by energy expended by individual constituents \cite{Marchetti2013}. 

A fruitful connection between active matter and biology is the widely accepted use of active nematic theory to model the interplay between cell shapes and tissue dynamics  \cite{Doostmohammadi2018,Balasubramaniam2022}. Nematic systems consist of elongated constituents that exhibit orientational order with no preferred direction within this orientation, i.e. they are head-tail symmetric. As such, the orientation of a cells long axis is a nematic object, and the average local direction of cell elongation can be thought of as a nematic field. The nematic field is coupled to the velocity field, with the energy expenditure of individual cells driving a rich variety of out of equilibrium behavior \cite{Dombrowski2004,Opathalage2019}. A particularly fertile line of study within confluent tissues is the formation, dynamics and properties of topological defects in the nematic field \cite{Giomi2013,Giomi2014}, as they have been found to mediate important homeostatic and morphogenetic processes \cite{Balasubramaniam2022}. 

\begin{figure}[!h]
    \begin{center}
        \includegraphics[width=100mm]{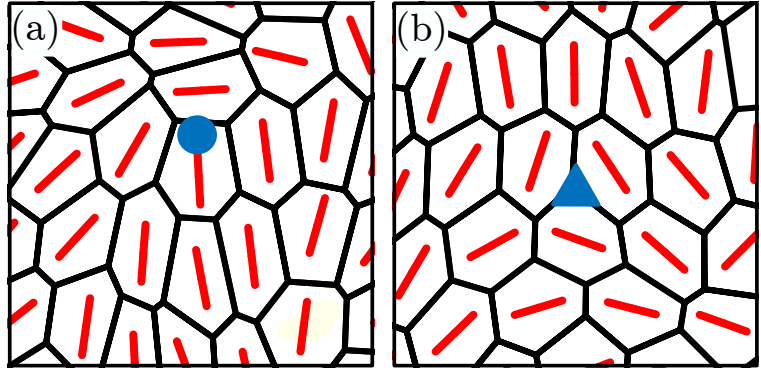}
    \end{center}
\caption{{\bf Topological defects in confluent tissues.} Examples of (a) comet-shaped $+1/2$ and (b) trefoil-shaped $-1/2$ defects in a confluent cell layer with the orientation of the long axis of each cell plotted in red.}
\label{fig1}
\end{figure}

Topological defects are singularities in the nematic field, points where its orientation does not vary smoothly but is discontinuous. In active nematic systems, two types of defects are typically found: comet-shaped singularities, known as $+1/2$ defects (Fig.\,\ref{fig1}a), and trefoil-shaped singularities, known as $-1/2$ defects (Fig.\,\ref{fig1}b). It is these nematic defects that are being highlighted as having a functional role in an increasing number of biological processes. Comet-shaped $+1/2$ defects have been found to trigger cell extrusion in epithelial layers \cite{Saw2017}, control the collective dynamics of confluent layers of neural progenitor cells \cite{Kawaguchi2017}, and have been highlighted as organisation centres during {\it Hydra} morphogenesis \cite{Maroudas-Sacks2021}. These $+1/2$ defects also mediate processes in densely packed bacterial systems, triggering the formation of fruiting bodies in {\it Myxococcus Xanthus} colonies \cite{Copenhagen2021}, as well as facilitating collective motion in {\it Pseudomonas aeruginosa} \cite{Meacock2021} and {\it E. Coli} colonies \cite{Doostmohammadi2016}. On the other hand, $-1/2$ defects have been associated with controlling areas of cell depletion in bacterial colonies \cite{Copenhagen2021}. We have also recently shown that active nematic defects can arise in confluent cell layers with no inherently nematic active forces \cite{Killeen2022}. Due to the prevalence and functional role of $\pm 1/2$ in many systems, the efficient detection and characterisation of topological defects in confluent tissues, and cellular systems generally, is of fundamental interest to both biology and physics.

While defect detection algorithms exist, their application to imaging data often requires a sophisticated understanding of the underlying physics. Current algorithms entail locating degenerate points in the nematic field followed by inspecting how the orientation of the nematic field changes around this point \cite{deGennes1974}, usually by calculating a quantity known as the winding number at this point. This can be effective in systems where the nematic field is well defined across the domain \cite{Decamp2015,Ellis2018} including tissues where cells are elongated or rod-like, such as spindle shaped fibroblasts \cite{Duclos2017}. However, this method is not suited to systems where the nematic field is not well-defined everywhere, as degenerate points in the field could just be areas of low order and do not necessarily indicate the existence of a defect. This is often the case in confluent tissues such as epithelial layers, where the cells are not rod-like and can be nearly isotropic in shape at times. Previous work studying defects in these systems searched for defects by calculating the winding number on a pre-defined grid of points in the nematic field \cite{Killeen2022,Saw2017,Balasubramaniam2021,Armengol-Collado2022}. This method required thousands of defects to be detected to adequately discern the average properties of defects, such as tissue stress and velocity fields, which are often the target of experimental studies. The necessity of such large amounts of data suggests the method of defect detection is inefficient and imprecise, which begs the question as to whether better methods of detection are possible for these systems.  

One possibility is to utilize machine learning to improve defect detection. Machine learning methods are being exploited in an increasing variety of applications in active nematic systems \cite{Hedlund2022,Colen2021,Zhou2021}. They have also been used to detect topological defects in various systems \cite{Beach2018,Walters2019,Minor2020}, including cellular systems \cite{Wenzel2021}. This previous study identified degenerate points in the nematic field of a cellular layer and then used a feed-forward, fully-connected neural network to perform a binary classification by labelling the points as either $+1/2$ or $-1/2$ defects. However, as previously discussed, this method is less applicable to experimental cellular systems where the nematic field is not well-defined everywhere, and points with low nematic order do not necessarily indicate the existence of a defect. Additionally, this study did not demonstrate that its machine learning method outperformed existing techniques for detection. As such, there is still a need to develop a machine learning method that can outperform current techniques and be readily utilized in an experimental setting.

Here, we address this problem by developing a methodology to detect nematic defects in confluent tissues using a convolutional neural network (CNN). We design the method such that it is well suited for use in systems that currently lack effective detection techniques, is user-friendly and readily implementable on experimental data. In contrast to previous work, we show that it outperforms current detection techniques and further demonstrate its efficacy by finding the mean velocity field around $+1/2$ defects and comparing this to defects detected using the winding number method, highlighting the improvement in capturing properties of topological defects with limited data.

\section*{Materials and methods}

\subsection*{Acquisition of test and training data}
For our method to be useful, it needs to be suitable for use on experimental data. For this reason, it takes as its input the $x$ and $y$ coordinates of each cells centre of mass and the orientation of the long axis of each cell, both of which can be readily acquired using standard segmentation software \cite{Schindelin2012}. However, a large amount of data is required to adequately train and test the model. Here, to obtain sufficient amounts of data, we train and test our model using data from a numerical model of a confluent cell layer: the active vertex model (AVM) \cite{Sussman2017}. 

\begin{figure}[!h]
    \begin{center}
        \includegraphics[width=135mm]{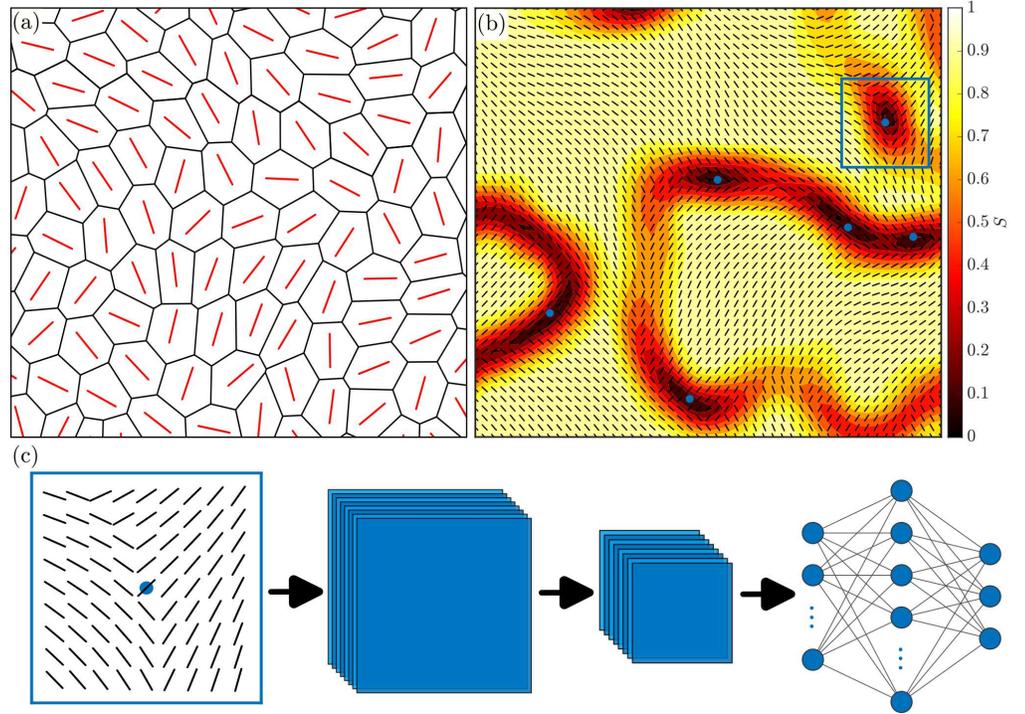}
    \end{center}
\caption{{\bf Defect identification and classification procedure.}
(a) Example of an active vertex model configuration. We find the $x$ and $y$ coordinates of each cells centre of mass, the orientation of each cells long axis is then plotted at these points. (b) This information is then interpolated to a finer grid to form the nematic field of the system, where the average local scalar nematic order parameter $S$ can be calculated at each grid point using a sliding window. Areas of low order ($S_{th}<0.15$) are identified as possible defect regions and the centres of mass of these regions are identified (blue dots). The nematic field around these points is then cropped to form a region of interest (ROI) (blue box). (c) These ROIs are then input into a machine learning model which classifies them as a $+1/2$ defect, a $-1/2$ defect or not a defect.}
\label{fig2}
\end{figure}

The AVM have been used extensively to study epithelial tissue dynamics \cite{Shaebani2020} and has been found to accurately replicate phenomena observed experimentally \cite{Petrolli2019,Henkes2020}. Moreover, data from an AVM is appropriate for developing our method as it represents the cell layer in a manner very similar to how they are represented once experimental images have been segmented (Fig.\,\ref{fig2}a). We implement the AVM in the same manner as our previous work \cite{Killeen2022}. Briefly, we represent the tissue as a confluent tiling of polygons, the degrees of freedom being the cell vertices. In the overdamped limit, these vertices move according to two types of forces: passive mechanical interactions between cells which arise due to gradients in an effective tissue energy function, and polar self-propulsive forces that model the motility of each cell. The effective tissue energy for a tissue containing $N$ cell is 
\begin{equation}
E = \Sigma^N_{a=1} (A_a-A_0)^2 + (P_a-P_0)^2 \ ,
\end{equation}
where $A_a$ and $P_a$ are the areas and perimeters of cell $a$ respectively, with $A_0$ and $P_0$ being the target area and perimeter for each cell. The first term encodes the incompressibility of each cell and the cell layers resistance to height fluctuations. The second term in the energy function encodes the competition between cortical tension and cell-cell adhesion. The force on vertex $i$ due to mechanical interactions is then $\mathbf{F}_i=-\nabla_{i}E$.  Self-propulsion is modeled by each cell generating a polar force of magnitude $f_0$, that acts along polarity vector $\hat{\bn}_a=(\cos{\theta_a},\sin{\theta_a})$. The self-propulsion force on each vertex is then the average self-propulsion of the three cells that neighbour vertex $i$, $\mathbf{f}_{i} = \frac{f_0}{3}\sum_{a \in \mathcal{N}(i)}\hat{\bn}_a$, where $\mathcal{N}(i)$ denotes the list of cells that share vertex $i$. Each vertex then moves according to
\begin{equation}
    \frac{d\mathbf{r}_{i}}{dt}=\frac{1}{\zeta}(\mathbf{F}_i + \mathbf{f}_{i}) \ ,
    \label{eq:ch4vertEOM}
\end{equation}
where $\zeta$ is the damping coefficient. The polarity vector of each cell undergoes rotational diffusion according to
\begin{equation}
    \frac{d\theta_a}{dt} = \sqrt{2D_r}\xi_a(t) \ ,
    \label{eq:ch4polEOM}
\end{equation}
where $\xi_a(t)$ is a white noise process with zero mean and unit variance and $D_r$ is the rotational diffusion coefficient. For a complete description of the AVM implementation please see \cite{Killeen2022}. Using the positions of the cell vertices, we find the centre of mass and long axis orientation of each cell, which we input into our model.  Details of parameter values used can be found in \nameref{S1_App}.

\subsection*{Identifying inputs to machine learning model}
To process our input such that it is in a form that a machine learning model can use to classify $\pm 1/2$ defects, we first identify `regions of interest' (ROIs) within the domain. As defects by definition occur in regions with low nematic order, we identify these areas as ROIs. To do this, we interpolate our cell orientation data to a fine grid and average each point over a sliding window to smooth out the data and create a nematic field (Fig.\,\ref{fig2}b). At each grid point we then calculate the scalar nematic order parameter $S$, which is defined as the largest eigenvalue of the nematic tensor $\mathbf{Q} = \langle 2\hat{u}_m\hat{u}_n - \delta_{mn} \rangle$, where $\mathbf{\hat{u}}=(\cos{\theta},\sin{\theta})$ with $\theta$ being the orientation of the field,  $\delta_{mn}$ is the kronecker delta and $\langle\cdot\rangle$ represents a spatial average over a sliding window. $S$ takes a value of one if the local nematic field is perfectly aligned and 0 if the local field is isotropic. As we seek areas of low order, we identify contiguous areas in our domain where $S$ is below a threshold value $S_{th}=0.15$. We then take the centres of mass of these areas to be the centres of our ROIs, cropping the field in a square around these points (Fig.\,\ref{fig2}b). We choose the value of $S_{th}$ such that it is high enough to capture all disordered regions that may contain defects, but low enough such that these regions are distinct and do not coalesce. We size our ROIs such that they contain 5-7 cells, large enough so as to capture the core of the defect but small enough to isolate the defects and avoid capturing multiple defects in a single ROI. We then use our ROIs as inputs into a machine learning model that classifies them as containing a $+1/2$ defect, $-1/2$ defect or neither (Fig.\,\ref{fig2}c). Details of parameter values used for pre-processing the data can be found in \nameref{S1_App}.
As the position of potential defect locations (the centres of our ROIs) can be located anywhere in the system domain, our method is effectively off-lattice in its detection, although they still lie on a fine grid. This contrasts with previous work detecting defects in epithelial cell layers \cite{Saw2017,Balasubramaniam2021}, which can only detect defects at predefined locations on a coarse-grained lattice.

\subsection*{Model architecture and training}
We use a CNN to classify our ROIs. A schematic of the architecture can be seen in Fig.\,\ref{fig3}c. We use two convolutional layers, each detecting 32 features. Due to the size of our ROIs, we do not use any max pooling layers after these convolutional layers. We then follow these convolutional layers with an additional fully-connected layer of 100 artificial neurons before our output layer of three neurons, representing our three possible outputs, or classes, of a $+1/2$ defect, $-1/2$ defect or no defect. Having a third output of no defect is key here and what makes our method particularly well suited to epithelial tissues. As the cells in our tissue do not have a well-defined long axis, neighbouring cells are not always nematically aligned and there can be regions with low nematic order that do not necessarily contain nematic defects. Including an option for our CNN to classify an ROI as having no defect accounts for this possibility. The output of our convolutional and fully-connected layers are rectified linear units, while the output layer is softmax \cite{Nielsen2015}.

To generate training and testing data we manually classify 5000 ROIs, using 4500 to train our model and saving 500 for testing. To enlarge our training data, we generate three new copies of each training ROI by rotating each one by angles $-\pi/2$, $\pi$ and $\pi/2$. Also, as the type of defect is invariant under reflections, we double this enlarged training data by reflecting each ROI about its centreline, leading to 36000 training inputs. We do not enlarge our testing data set.

We train our model by minimising the cross-entropy cost function, defined as
\begin{equation}
C = -\sum_{i=1}^{N}\sum^3_{c=1}y_{i,c}\log{p_{i,c}} \ ,
\end{equation}
where $N$ is the number of items in each batch of training data, $y_{i,c}$ is the correct label (0 or 1) for class $c$ of the $i^{th}$ ROI and $p_{i,c}$ is the probability calculated by the model that the $i^{th}$ ROI belongs to class $c$. We minimise $C$ using a stochastic gradient descent algorithm with a batch size of $N= 64$ ROIs. We train over 20 epochs with a learning rate of 0.05 for the first ten epochs and 0.005 for the following ten, initializing our weights using a Glorot normal distribution \cite{Glorot2010}. For each epoch, a random 10\% of our training images are held back for validation and used at the end of the epoch to assess the accuracy of our model. Training for more than 20 epochs did not lead to any appreciable improvements in validation accuracy. A complete list of parameter values used can be found in  \nameref{S1_App}.

\subsection*{Winding number calculation and comparison}
Manually labelling the ROIs allows a direct comparison between our method and the current standard technique used in defect detection, calculation of the winding number, to be drawn, as we can also classify each ROI by calculating its winding number. We can then find the accuracy of both methods when compared against our manually labelled ROIs, our `ground truth'. Previous work on applying machine learning to detect nematic defects in tissues has used the winding number as the ground truth \cite{Wenzel2021}, thereby making it impossible to determine if the machine learning method is superior to current techniques.

The winding number is the amount the nematic field rotates as a closed loop is traversed around the centre of the defect \cite{Huterer2005}. The $\pm1/2$ defects found in nematic systems are so called because the nematic field rotates by half a full rotation, or $\pi$ radians, around the loop (\nameref{S1_Fig}). The sign of the defects depends on whether the rotation of the nematic field is in the same direction as the direction in which the loop is being traversed. If the nematic field rotates clockwise as the loop is traversed in a clockwise direction, the defect is positive; if it rotates anti-clockwise, it is negative. We classify each ROI by finding the winding number on the fine grid around the edge of the ROI.

\section*{Results}
\subsection*{Machine learning model outperforms winding number classification}
\begin{figure}[!b]
    \begin{center}
        \includegraphics[width=135mm]{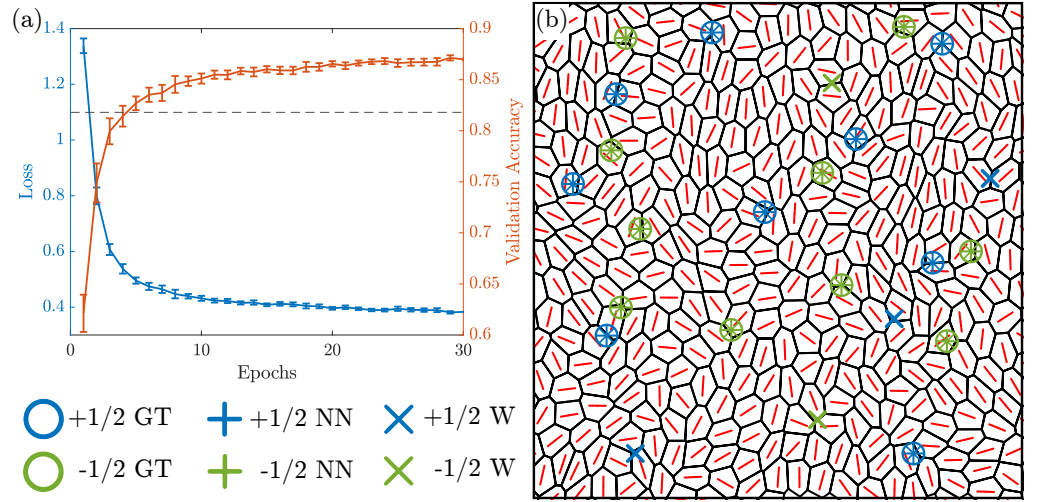}
    \end{center}
\caption{{\bf Machine learning model outperforms winding number classifier.}
(a) The loss and validation accuracy of the neural network as it is trained. The dashed line represents the accuracy of the winding number classification on the training data (0.812). Error bars represent the standard error in the mean over 50 realisations. (b) An example domain with defects detected using each method: our ground truth (GT), neural network (NN) and winding number (W). }
\label{fig3}
\end{figure}

The mean performance of our CNN model with each training epoch can be seen in Fig.\,\ref{fig3}a. After training, our model clearly outperforms the winding number for overall classification accuracy on the training data set, defined as the percentage of correct predictions. However, these are ROIs that our model is being trained on, meaning it has `seen' them before in previous training epochs. The real utility of our method depends on its ability to classify ROIs it has not seen before, which we test using the 500 ROIs in our test data set. Here our model is again more accurate than the winding number, achieving an accuracy of 84.0\% compared to the winding numbers 76.6\%, demonstrating that our method outperforms the current most widely used technique.

Defect detection techniques can often be sensitive to the window size used to detect them. If our trained model is to be readily usable on experimental data, it should achieve accurate results over a range of window sizes. To investigate this, we assess the accuracy of our trained model and the winding number method in classifying the test data at different grid sizes (See \nameref{S1_App}). As our model takes as input a $9\times9$ grid of points, changing the grid size is akin to changing the ROI size. We find that, over a range of grid sizes, our method outperforms the winding number method, demonstrating its robustness in classifying defects even when the ROI size is not well tuned to the size of defects in the system (See Fig\,1 in \nameref{S1_App}).

As an example of the defects detected using each method, we look at an example domain from our AVM containing ROIs from our test data set (Fig.\,\ref{fig3}b). In line with Fig.\,\ref{fig3}a, both techniques show good agreement with manually labelled defects although the winding number appears to detect more false positives than the neural network. To assess this further and properly delineate the efficacy of both methods, we break down their performance for each class in Table \ref{table1}. We calculate the precision, sensitivity and F1 score of each method. The precision is defined as 
\begin{equation}
\textrm{P} = \frac{\textrm{True Positive}}{\textrm{True Positive + False Positive}} \ ,
\end{equation}
and determines what proportion of detections are correct. It quantifies how detrimental false positives are to performance. Sensitivity, on the other hand, examines the role of false negatives in performance. It is defined as 
\begin{equation}
\textrm{S} = \frac{\textrm{True Positive}}{\textrm{True Positive + False Negative}} \ ,
\end{equation}
and establishes what proportion of true defects are actually identified. The F1 score, defined as
\begin{equation}
\textrm{F1} = \frac{2\textrm{PS}}{\textrm{P + S}} \ ,
\end{equation}
is a weighted average of the precision and sensitivity and so is a broader metric of performance.

\begin{table}[!ht]
\begin{adjustwidth}{-2.25in}{0in} 
\centering
\caption{
{\bf Performance of defect detection methods on test data.} Precision (P), sensitivity (S) and F1 score for each class, as well as an average over all classes, weighted by the size of that class.}
\begin{tabular}{|c+ccc+ccc+ccc+ccc|}
\hline
 \multirow{2}{*}{} & \multicolumn{3}{c+}{\bf $\mathbf{+1/2}$} & \multicolumn{3}{c+}{\bf No defect} & \multicolumn{3}{c+}{\bf $\mathbf{-1/2}$} & \multicolumn{3}{c|}{\bf Total}\\ 
 & P & S & F1 & P & S & F1 & P & S & F1 & P & S & F1\\ \thickhline
Neural Network & 0.786 & 0.967 & 0.867 & 0.932 & 0.663 & 0.775 & 0.818 & 0.964 & 0.885 & 0.856 & 0.840 & 0.834\\
Winding Number & 0.729 & 0.974 & 0.834 & 0.960 & 0.457 & 0.619 & 0.707 & 1.000 & 0.828 & 0.819 & 0.766 & 0.743\\ \hline
\end{tabular}
\begin{flushleft} 
\end{flushleft}
\label{table1}
\end{adjustwidth}
\end{table}

Both methods display a similar pattern of having a higher sensitivity than precision for both defect categories, but a higher precision than sensitivity when no defect is present. Additionally, both methods exhibit a lower F1 score when no defect is present, a reflection of the larger differential between precision and sensitivity scores. Errors in both methods, therefore, primarily come from falsely detecting defects, as opposed to missing defects that should be detected. This information is lost when looking just at the weighted average values across all classes, which give more comparable precision and sensitivity scores. 

Where the two methods differ is in our CNN model having consistently higher F1 for each category. This is driven by its higher precision in each defect class and higher sensitivity when no defect is present. The winding number, however, is slightly more sensitive to detecting defects when they are present. Taken together, these results show that the improved performance of our neural network compared to the winding number primarily stems from it detecting fewer false positive defects. We point out here that it could be argued that precision and sensitivity should not be weighted equally, as they are in the F1 score, and that there may be scenarios where ensuring detecting as many defects as possible is more important than minimising detecting defects that are not there. However, we now show that, while the winding number may detect a slightly higher proportion of defects, the higher overall performance of our model can manifest itself in a wider improvement to experimental results. 

\subsection*{Superior performance leads to improved capturing of defect properties}

While results thus far point to the effectiveness of our model, to show that this realizes itself in tangible improvements to wider results, we look at the ability of our model, and the winding number, to ascertain the properties of defects. Experimental studies often seek not only to detect defects but examine tissue properties around them \cite{Saw2017,Balasubramaniam2021}. To this end, we calculate the average velocity field around $+1/2$ defects detected using each method. This observable is particularly pertinent as one can infer global system properties from the velocity of $+1/2$ defects. The velocity direction indicates whether the system is behaving as an extensile (the net force on cells is pushing out along its long axis) or contractile (pulling in along its long axis) nematic, with tail-to-head motion indicating extensile forces and head-to-tail motion indicating contractile \cite{Giomi2014}. Epithelial layers have been shown to exhibit both forward and backward motion in experiments \cite{Balasubramaniam2021}. It is therefore valuable to be able to distinguish defect motion accurately and efficiently.

Previous work using this AVM has determined that $+1/2$ defects move in a tail-to-head direction, indicating extensile behavior, in this system \cite{Killeen2022}. However, obtaining the characteristic extensile flow field required averaging the velocity field over many simulations and several thousand defects. While this was achievable in a numerical model, time and cost constraints could make the requirements of such vast amounts of data to understand the properties of these defects prohibitive in an experimental setting. Due to the difficulty in collecting experimental data, it is therefore crucial that defect properties can be discerned using a minimal amount of information. 

\begin{figure}[!b]
    \begin{center}
        \includegraphics[width=135mm]{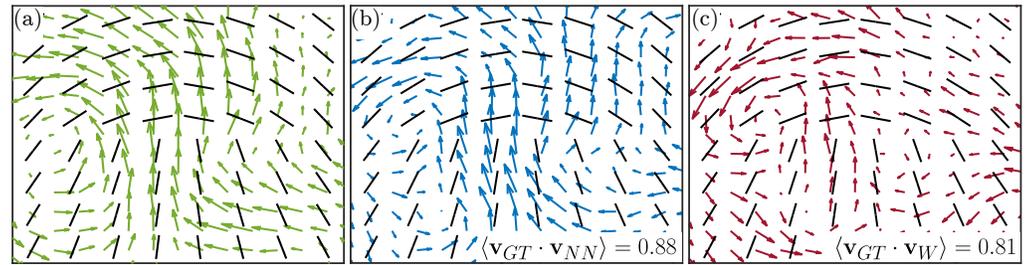}
    \end{center}
\caption{{\bf Less data is needed to characterise defect properties.}
Average velocity fields around $+1/2$ defects for (a) manually labelled defects, (b) defects detected using the neural network and (c) defects detected using the winding number. The single point correlation function ($\langle \mathbf{v}_{GT}\cdot\mathbf{v}_{\Box}\rangle$) between the ground truth (GT) field and the neural network (NN) and winding number (W) fields is also shown.}
\label{fig4}
\end{figure}

\begin{figure}[!b]
    \begin{center}
        \includegraphics[width=135mm]{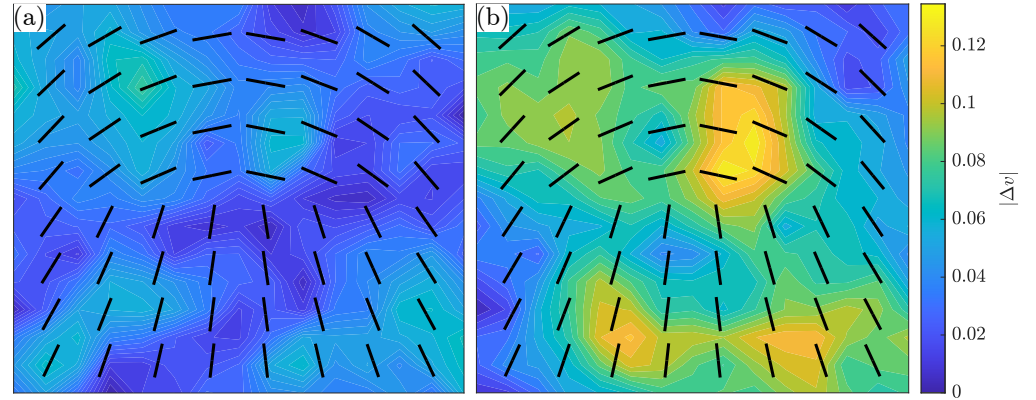}
    \end{center}
\caption{{\bf Winding number velocity field exhibits larger error than machine learning flow field.}
Difference in velocity magnitudes between average velocity fields for manually labelled defects and (a) defects detected using the neural network and (b) defects detected using the winding number.}
\label{fig5}
\end{figure}

The average velocity fields for manually labelled, winding number detected and neural network detected defects, using 150 $+1/2$ defects detected from the test data set, can be seen in Fig.\,\ref{fig4}. 150 defects were used as this was number of $+1/2$ defects manually labelled in the test data set and so the largest number we could use to compare the different methods. The manually labelled defects demonstrate the clearest tail-to-head, vortical flow fields characteristic of extensile systems \cite{Giomi2014}. The flow field around CNN detected defects clearly show better agreement with the manually labelled flow field than the flow field found using the winding number, reflected in the higher correlation between the two fields. The improvement between the two methods is even more stark when looking at the difference in velocity magnitudes between each method and the manually labelled flow field (Fig.\,\ref{fig5}). This illustrates the impact of the reduced performance, particularly the reduced precision, of the winding number and confirms the primacy of our model in detecting `better' defects, as the anticipated mean-field behavior is clearer.

Additionally, we compared the velocity field around defects detected using our model and defects detected on the same data using the `on-lattice' winding number method used previously \cite{Killeen2022} (\nameref{S2_Fig}). The difference between the two is even starker than between our model and the off-lattice winding number method used in the present study. As well as highlighting the improvement using an off-lattice method can bring, it further underlines the benefits of our model compared to techniques used currently. 

\section*{Discussion}
In this study, we have developed a new method for detecting nematic defects in confluent tissues which, crucially, is readily implementable on experimental data. Our model can therefore aid in the expanding study of characterising cellular layers as active nematic systems, as active nematic defects are increasingly found to play functional roles in these systems \cite{Doostmohammadi2018}. Importantly, we demonstrate that our method displays superior performance to the current standard use of the winding number in detecting defects and in capturing the mean-field properties of these defects. This reduces the amount of data required to obtain these properties, potentially improving experimental data interpretation.

Interestingly, although the overall performance of our model is better, the winding number is slightly more sensitive to detecting defects. This means there could be applications where using the winding number would be more suitable, if the cost of missing a defect in the domain is very high. However, the improved performance in finding mean defect flow fields demonstrate that, in practice, the increase in overall performance of our model makes it more advantageous. This improved performance is likely due to the winding number only using information around the edge of the ROI, where as our CNN can take advantage of spatial information and correlations across the whole region.

In contrast to previous studies on using machine learning to detect nematic defects \cite{Wenzel2021}, our method is specifically designed for noisy experimental systems where the nematic field may not be well defined everywhere, and consequently low nematic order may not guarantee the presence of a defect. However, we anticipate our method will work well with any system whose nematic field can be easily interpolated to a 2D grid. Indeed, applying our method to active nematic systems, such as microtubule systems \cite{Sanchez2012}, would be an interesting future application of our technique. Another interesting future avenue of research includes extending the model to detect integer $+1$ defects, such as spiral or aster shaped singularities, as these have been engineered to arise in cellular systems \cite{Turiv2020,Endresen2021}, and have also been linked to morphogenetic processes \cite{Maroudas-Sacks2021, Guillamat2022}.

The trained CNN model along with training data and Python scripts to detect defects can be found at \url{https://github.com/KilleenA/ML_DefectDetection}.

\section*{Supporting information}

\paragraph*{S1 Appendix}
\label{S1_App}
{\bf Parameter values used in AVM, hyperparameters used in machine learning model and details of the grid size study.} 
\paragraph*{S1 Fig.}
\label{S1_Fig}
{\bf Calculating a defects winding number.} (a) Example of a trefoil-shaped $-1/2$ defects in a confluent cell layer with the orientation of the long axis of each cell plotted in red. (b) Characterising this defect by its winding number. As a closed loop is traversed around the $-1/2$ defect, the orientation of the cells rotate by $\pi$ radians (half a full rotation), hence the defect is half-integer. The sign of the defect is negative as the cells rotate in the opposite direction to the direction of travel around the loop
\begin{figure*}[!h]
    \begin{center}
        \includegraphics[width=100mm]{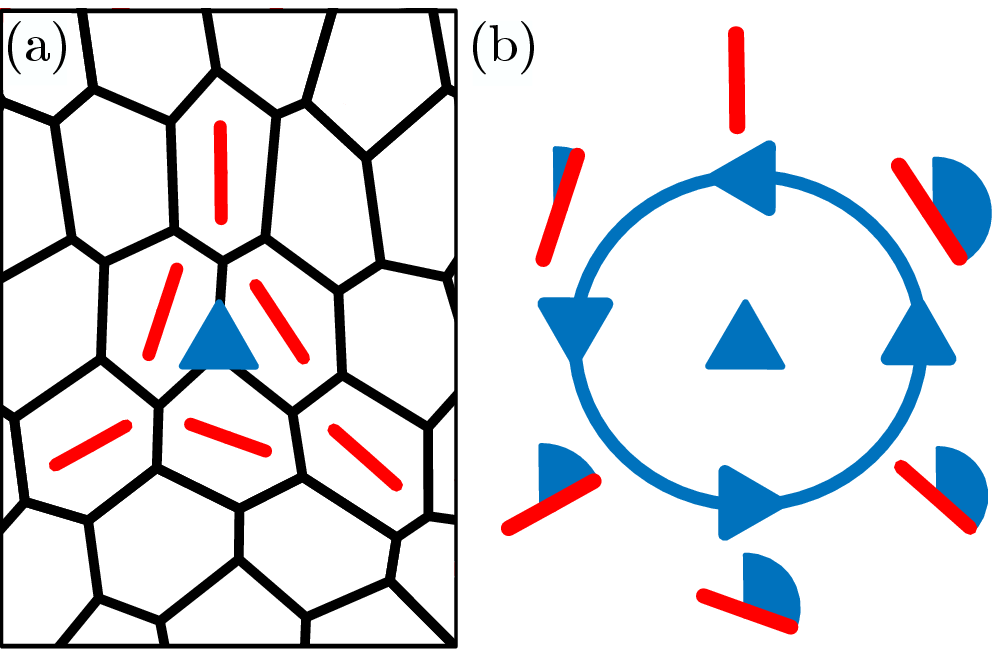}
    \end{center}
\end{figure*}

\paragraph*{S2 Fig.}
\label{S2_Fig}
{\bf Flow fields around $+1/2$ defects.} Mean tissue velocity fields around 150 $+1/2$ defects detected using (a) our CNN model and (b) the winding number at predefined points in the domain, used previously \cite{Killeen2022}.
\begin{figure*}[!h]
    \begin{center}
        \includegraphics[width=115mm]{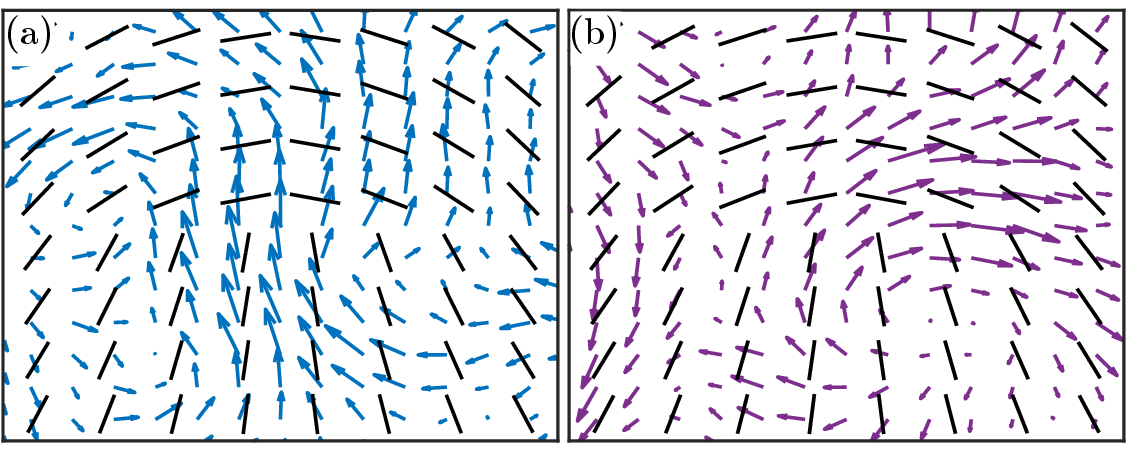}
    \end{center}
\end{figure*}

\section*{Acknowledgments}
We acknowledge the High Throughput Computing service provided by Imperial College Research Computing Service. AK was supported by the EPSRC Centre for Doctoral Training in Fluid Dynamics Across Scales (Grant EP/L016230/1).


\bibliography{references}

\begin{thebibliography}{10}

\bibitem{Poujade2007}
Poujade M, Grasland-Mongrain E, Hertzog A, Jouanneau J, Chavrier P, Ladoux B,
  et~al.
\newblock Collective migration of an epithelial monolayer in response to a
  model wound.
\newblock Proceedings of the National Academy of Sciences.
  2007;104(41):15988--15993.
\newblock doi:{10.1073/pnas.0705062104}.

\bibitem{Friedl2009}
Friedl P, Gilmour D.
\newblock {Collective cell migration in morphogenesis, regeneration and
  cancer}.
\newblock Nature Reviews Molecular Cell Biology. 2009;10(7):445--457.
\newblock doi:{10.1038/nrm2720}.

\bibitem{Weijer2009}
Weijer CJ.
\newblock {Collective cell migration in development}.
\newblock Journal of Cell Science. 2009;122(18):3215--3223.
\newblock doi:{10.1242/jcs.036517}.

\bibitem{Alert2020}
Alert R, Trepat X.
\newblock {Physical Models of Collective Cell Migration}.
\newblock Annual Review of Condensed Matter Physics. 2020;11(1):77--101.
\newblock doi:{10.1146/annurev-conmatphys-031218-013516}.

\bibitem{Marchetti2013}
Marchetti MC, Joanny JF, Ramaswamy S, Liverpool TB, Prost J, Rao M, et~al.
\newblock {Hydrodynamics of soft active matter}.
\newblock Reviews of Modern Physics. 2013;85(3):1143--1189.
\newblock doi:{10.1103/RevModPhys.85.1143}.

\bibitem{Doostmohammadi2018}
Doostmohammadi A, Ign{\'{e}}s-Mullol J, Yeomans JM, Sagu{\'{e}}s F.
\newblock {Active nematics}.
\newblock Nature Communications. 2018;9(1):3246.
\newblock doi:{10.1038/s41467-018-05666-8}.

\bibitem{Balasubramaniam2022}
Balasubramaniam L, Mège RM, Ladoux B.
\newblock Active nematics across scales from cytoskeleton organization to
  tissue morphogenesis.
\newblock Current Opinion in Genetics \& Development. 2022;73:101897.
\newblock doi:{https://doi.org/10.1016/j.gde.2021.101897}.

\bibitem{Dombrowski2004}
Dombrowski C, Cisneros L, Chatkaew S, Goldstein RE, Kessler JO.
\newblock {Self-concentration and large-scale coherence in bacterial dynamics}.
\newblock Physical Review Letters. 2004;93(9):098103.
\newblock doi:{10.1103/PhysRevLett.93.098103}.

\bibitem{Opathalage2019}
Opathalage A, Norton MM, Juniper MPN, Langeslay B, Aghvami SA, Fraden S, et~al.
\newblock {Self-organized dynamics and the transition to turbulence of confined
  active nematics}.
\newblock Proceedings of the National Academy of Sciences of the United States
  of America. 2019;116(11):4788--4797.
\newblock doi:{10.1073/pnas.1816733116}.

\bibitem{Giomi2013}
Giomi L, Bowick MJ, Ma X, Marchetti MC.
\newblock {Defect annihilation and proliferation in active Nematics}.
\newblock Physical Review Letters. 2013;110(22):228101.
\newblock doi:{10.1103/PhysRevLett.110.228101}.

\bibitem{Giomi2014}
Giomi L, Bowick MJ, Mishra P, Sknepnek R, Cristina~Marchetti M.
\newblock Defect dynamics in active nematics.
\newblock Philosophical Transactions of the Royal Society A: Mathematical,
  Physical and Engineering Sciences. 2014;372(2029):20130365.
\newblock doi:{10.1098/rsta.2013.0365}.

\bibitem{Saw2017}
Saw TB, Doostmohammadi A, Nier V, Kocgozlu L, Thampi S, Toyama Y, et~al.
\newblock Topological defects in epithelia govern cell death and extrusion.
\newblock Nature. 2017;544:212--216.
\newblock doi:{10.1038/nature21718}.

\bibitem{Kawaguchi2017}
Kawaguchi K, Kageyama R, Sano M.
\newblock {Topological defects control collective dynamics in neural progenitor
  cell cultures}.
\newblock Nature. 2017;545(7654):327--331.
\newblock doi:{10.1038/nature22321}.

\bibitem{Maroudas-Sacks2021}
Maroudas-Sacks Y, Garion L, Shani-Zerbib L, Livshits A, Braun E, Keren K.
\newblock Topological defects in the nematic order of actin fibres as
  organization centres of Hydra morphogenesis.
\newblock Nature Physics. 2021;17(2):251--259.
\newblock doi:{10.1038/s41567-020-01083-1}.

\bibitem{Copenhagen2021}
Copenhagen K, Alert R, Wingreen NS, Shaevitz JW.
\newblock Topological defects promote layer formation in Myxococcus xanthus
  colonies.
\newblock Nature Physics. 2021;17(2):211--215.
\newblock doi:{10.1038/s41567-020-01056-4}.

\bibitem{Meacock2021}
Meacock OJ, Doostmohammadi A, Foster KR, Yeomans JM, Durham WM.
\newblock Bacteria solve the problem of crowding by moving slowly.
\newblock Nature Physics. 2021;17(2):205--210.
\newblock doi:{10.1038/s41567-020-01070-6}.

\bibitem{Doostmohammadi2016}
Doostmohammadi A, Thampi SP, Yeomans JM.
\newblock {Defect-Mediated Morphologies in Growing Cell Colonies}.
\newblock Physical Review Letters. 2016;117(4):048102.
\newblock doi:{10.1103/PhysRevLett.117.048102}.

\bibitem{Killeen2022}
Killeen A, Bertrand T, Lee CF.
\newblock Polar Fluctuations Lead to Extensile Nematic Behavior in Confluent
  Tissues.
\newblock Phys Rev Lett. 2022;128:078001.
\newblock doi:{10.1103/PhysRevLett.128.078001}.

\bibitem{deGennes1974}
de~Gennes PG.
\newblock The physics of liquid crystals.
\newblock Oxford: Oxford University Press; 1974.

\bibitem{Decamp2015}
DeCamp SJ, Redner GS, Baskaran A, Hagan MF, Dogic Z.
\newblock {Orientational order of motile defects in active nematics}.
\newblock Nature Materials. 2015;14(11):1110--1115.

\bibitem{Ellis2018}
Ellis PW, Pearce DJG, Chang YW, Goldsztein G, Giomi L, Fernandez-Nieves A.
\newblock {Curvature-induced defect unbinding and dynamics in active nematic
  toroids}.
\newblock Nature Physics. 2018;14(1):85--90.
\newblock doi:{10.1038/nphys4276}.

\bibitem{Duclos2017}
Duclos G, Erlenk{\"{a}}mper C, Joanny JF, Silberzan P.
\newblock {Topological defects in confined populations of spindle-shaped
  cells}.
\newblock Nature Physics. 2017;13(1):58--62.
\newblock doi:{10.1038/nphys3876}.

\bibitem{Balasubramaniam2021}
Balasubramaniam L, Doostmohammadi A, Saw TB, Narayana GHNS, Mueller R, Dang T,
  et~al.
\newblock Investigating the nature of active forces in tissues reveals how
  contractile cells can form extensile monolayers.
\newblock Nature Materials. 2021;20:1156--1166.
\newblock doi:{10.1038/s41563-021-00919-2}.

\bibitem{Armengol-Collado2022}
Armengol-Collado JM, Carenza LN, Eckert J, Krommydas D, Giomi L. Epithelia are
  multiscale active liquid crystals; 2022.
\newblock Available from: \url{https://arxiv.org/abs/2202.00668}.

\bibitem{Hedlund2022}
Hedlund E, Hedlund K, Green A, Chowdhury R, Park CS, Maclennan JE, et~al.
\newblock Detection of islands and droplets on smectic films using machine
  learning.
\newblock Physics of Fluids. 2022;34(10):103608.
\newblock doi:{10.1063/5.0117358}.

\bibitem{Colen2021}
Colen J, Han M, Zhang R, Redford SA, Lemma LM, Morgan L.
\newblock {Machine learning active-nematic hydrodynamics}.
\newblock PNAS. 2021;118(10):e2016708118.
\newblock doi:{10.1073/pnas.2016708118/-/DCSupplemental.y}.

\bibitem{Zhou2021}
Zhou Z, Joshi C, Liu R, Norton MM, Lemma L, Dogic Z, et~al.
\newblock {Machine learning forecasting of active nematics}.
\newblock Soft Matter. 2021;17(3):738--747.
\newblock doi:{10.1039/D0SM01316A}.

\bibitem{Beach2018}
Beach MJS, Golubeva A, Melko RG.
\newblock {Machine learning vortices at the Kosterlitz-Thouless transition}.
\newblock Physical Review B. 2018;97(4):045207.
\newblock doi:{10.1103/PhysRevB.97.045207}.

\bibitem{Walters2019}
Walters M, Wei Q, Chen JZY.
\newblock {Machine learning topological defects of confined liquid crystals in
  two dimensions}.
\newblock Physical Review E. 2019;99(6):062701.
\newblock doi:{10.1103/PhysRevE.99.062701}.

\bibitem{Minor2020}
Minor EN, Howard SD, Green AAS, Glaser MA, Park CS, Clark NA.
\newblock {End-to-end machine learning for experimental physics: Using
  simulated data to train a neural network for object detection in video
  microscopy}.
\newblock Soft Matter. 2020;16(7):1751--1759.
\newblock doi:{10.1039/c9sm01979k}.

\bibitem{Wenzel2021}
Wenzel D, Nestler M, Reuther S, Simon M, Voigt A.
\newblock {Defects in Active Nematics-Algorithms for Identification and
  Tracking}.
\newblock Computational Methods in Applied Mathematics. 2021;21(3):683--692.
\newblock doi:{10.1515/cmam-2020-0021}.

\bibitem{Schindelin2012}
Schindelin J, Arganda-Carreras I, Frise E, Kaynig V, Longair M, Pietzsch T,
  et~al.
\newblock {Fiji: An open-source platform for biological-image analysis}.
\newblock Nature Methods. 2012;9(7):676--682.
\newblock doi:{10.1038/nmeth.2019}.

\bibitem{Sussman2017}
Sussman DM.
\newblock cellGPU: Massively parallel simulations of dynamic vertex models.
\newblock Computer Physics Communications. 2017;219:400--406.
\newblock doi:{https://doi.org/10.1016/j.cpc.2017.06.001}.

\bibitem{Shaebani2020}
Shaebani MR, Wysocki A, Winkler RG, Gompper G, Rieger H.
\newblock {Computational models for active matter}.
\newblock Nature Reviews Physics. 2020;2(4):181--199.
\newblock doi:{10.1038/s42254-020-0152-1}.

\bibitem{Petrolli2019}
Petrolli V, {Le Goff} M, Tadrous M, Martens K, Allier C, Mandula O, et~al.
\newblock {Confinement-Induced Transition between Wavelike Collective Cell
  Migration Modes}.
\newblock Physical Review Letters. 2019;122(16):168101.
\newblock doi:{10.1103/PhysRevLett.122.168101}.

\bibitem{Henkes2020}
Henkes S, Kostanjevec K, Collinson JM, Sknepnek R, Bertin E.
\newblock {Dense active matter model of motion patterns in confluent cell
  monolayers}.
\newblock Nature Communications. 2020;11(1):1405.
\newblock doi:{10.1038/s41467-020-15164-5}.

\bibitem{Nielsen2015}
Nielsen M.
\newblock {Neural Networks and Deep Learning}.
\newblock Determination Press; 2015.

\bibitem{Glorot2010}
Glorot X, Bengio Y.
\newblock Understanding the difficulty of training deep feedforward neural
  networks.
\newblock In: Teh YW, Titterington M, editors. Proceedings of the Thirteenth
  International Conference on Artificial Intelligence and Statistics. vol.~9 of
  Proceedings of Machine Learning Research. Chia Laguna Resort, Sardinia,
  Italy; 2010. p. 249--256.

\bibitem{Huterer2005}
Huterer D, Vachaspati T.
\newblock {Distribution of singularities in the cosmic microwave background
  polarization}.
\newblock Physical Review D - Particles, Fields, Gravitation and Cosmology.
  2005;72(4):043004.
\newblock doi:{10.1103/PhysRevD.72.043004}.

\bibitem{Sanchez2012}
Sanchez T, Chen DTN, Decamp SJ, Heymann M, Dogic Z.
\newblock {Spontaneous motion in hierarchically assembled active matter}.
\newblock Nature. 2012;491(7424):431--434.
\newblock doi:{10.1038/nature11591}.

\bibitem{Turiv2020}
Turiv T, Krieger J, Babakhanova G, Yu H, Shiyanovskii SV, Wei QH, et~al.
\newblock Topology control of human fibroblast cells monolayer by liquid
  crystal elastomer.
\newblock Science Advances. 2020;6(20):eaaz6485.
\newblock doi:{10.1126/sciadv.aaz6485}.

\bibitem{Endresen2021}
Endresen KD, Kim MS, Pittman M, Chen Y, Serra F.
\newblock {Topological defects of integer charge in cell monolayers}.
\newblock Soft Matter. 2021;17(24):5878--5887.
\newblock doi:{10.1039/D1SM00100K}.

\bibitem{Guillamat2022}
Guillamat P, Blanch-Mercader C, Pernollet G, Kruse K, Roux A.
\newblock {Integer topological defects organize stresses driving tissue
  morphogenesis}.
\newblock Nature Materials. 2022;21(5):588--597.
\newblock doi:{10.1038/s41563-022-01194-5}.

\end{thebibliography}
%
%
%

%
%
%
%

\end{document}


\title{Supplemental material for machine learning of topological defects in confluent tissues. }
\author{Andrew Killeen}
\email{a.killeen18@imperial.ac.uk}
\affiliation{Department of Bioengineering, Imperial College London, South Kensington Campus, London SW7 2AZ, U.K.}
\author{Thibault Bertrand}
\email{t.bertrand@imperial.ac.uk}
\affiliation{Department of Mathematics, Imperial College London, South Kensington Campus, London SW7 2AZ, U.K.}
\author{Chiu Fan Lee}
\email{c.lee@imperial.ac.uk}
\affiliation{Department of Bioengineering, Imperial College London, South Kensington Campus, London SW7 2AZ, U.K.}
\maketitle

\section{AVM implementation and parameters}
We simulate $N = 400$ cells in a fluid like state with parameters $A_0=1$, $P_0=3.7$, $f_0=0.5$, $\zeta=1$ and $D_r=1$. We numerically integrate the equations of motion using the Euler-Maruyama method with time step $\Delta t=0.01$. 

We initialize the simulation by arranging $N$ cells on a hexagonal lattice, with grid spacing $d=\sqrt{2/\sqrt{3}}$, in a periodic domain with dimensions $dN\times (\sqrt{2}/3)dN$. We then draw a Voronoi diagram from the seeded points to obtain the initial positions of the vertices. The choice of grid spacing gives edges of length $a = d/\sqrt{3}$ and ensures that all cells initially have unit area. This means the average cell area throughout the simulation $\bar{A}=1$ and we use $\sqrt{\bar{A}}$ as our unit length. To ensure different realisations of the system were independent, cells have random initial polarities and we integrate through at least $2\times10^3$ time units to initialise the system. 

\section{Machine learning model implementation and parameters}
To identify ROIs we interpolate our input data to a fine grid with grid spacing $\Delta x$ = $\Delta y$ = 0.2, where the average cell length is approximately 1 length unit. We then smooth the data by passing a sliding window of size $9\times9$ over the data. We use a window of the same size to calculate the scalar order parameter at each point. Our threshold value of $S$ for identifying ROIs $S_{th}=0.15$. Our ROIs are then also $9\times9$ in size, meaning the inputs to our model are $9\times9$ grids.

We implement our model in Python using the TensorFlow library. Our convolutional neural network (CNN) has two layers, both detecting 32 features, the first has feature detectors of size $6\times6$ and the second $3\times3$. Our 100 neuron fully-connected layer uses L2-regularisation with strength $\lambda=0.01$. We chose this architecture as it achieved the highest classification accuracy on the training data, although we note that our results are not sensitive to the particular architecture used and similar architectures achieved comparable, albeit slightly lower, accuracies. To avoid overfitting when training the model, we use dropout on the fully-connected layer, leaving out a random 50\% of the neurons in each training batch. All layers use rectified linear units as their output function with the exception of the output layer, which uses softmax. 

\section{Sensitivity to grid size}
To further evidence the suitability of our method for experimental data analysis, we examine the sensitivity of the model to the size of the fine grid to which we interpolate the input data. Different systems will have different sized cells and different sized defects relative to the size of these cells. So, while we tune our window size to the size of defects in our system, it may be less clear what the correct size is in experimental systems, so our model should be able to accommodate different grid sizes without it impacting performance. To investigate this, we assess the accuracy of our trained model and the winding number method in classifying the test data at different grid sizes. We do not retrain our model with input data at the new grid size, we use the original model where the weights were trained on data interpolated to the original grid size ($\Delta x=0.2$). To enable comparisons with our ground truth, for each ROI we take the coordinates of the centre and interpolate the cell data to a new grid size about this central point. As our model takes as its input a $9\times9$ grid, inputting data at a new grid size is the same as changing the size of our ROIs around the defect center. We look at a range of grid sizes from $\Delta x=0.1$, which gives an ROI with a window length approximately one cell across, to $\Delta x=0.8$, meaning our interpolated grid is of the same order as the typical cell length and the ROI window lengths are approximately four times larger than the typical defect size. When smoothing the interpolated field, we scale the size of our sliding window such that the area over which we average is approximately constant.

The classification accuracy as a function of grid size can be seen in Fig.\,1. With the exception of the smallest grid size, our CNN consistently outperforms the winding number. Along with being more accurate, our method is also less sensitive to the grid size than the winding number, whose performance drops sharply when the grid size varies from that which gives the highest accuracy, and is no better than random selection when the grid size is greater than 0.5. The CNN, however, is able to maintain a level of accuracy greater than, or similar to, the winding number's best performance even when the ROI is approximately four times larger than the size of the defect. This demonstrates the robustness of our approach and the ability of method to detect defects even when the parameters of the model are not perfectly tuned to the system.

\begin{figure}[!h]
    \begin{center}
        \includegraphics[width=100mm]{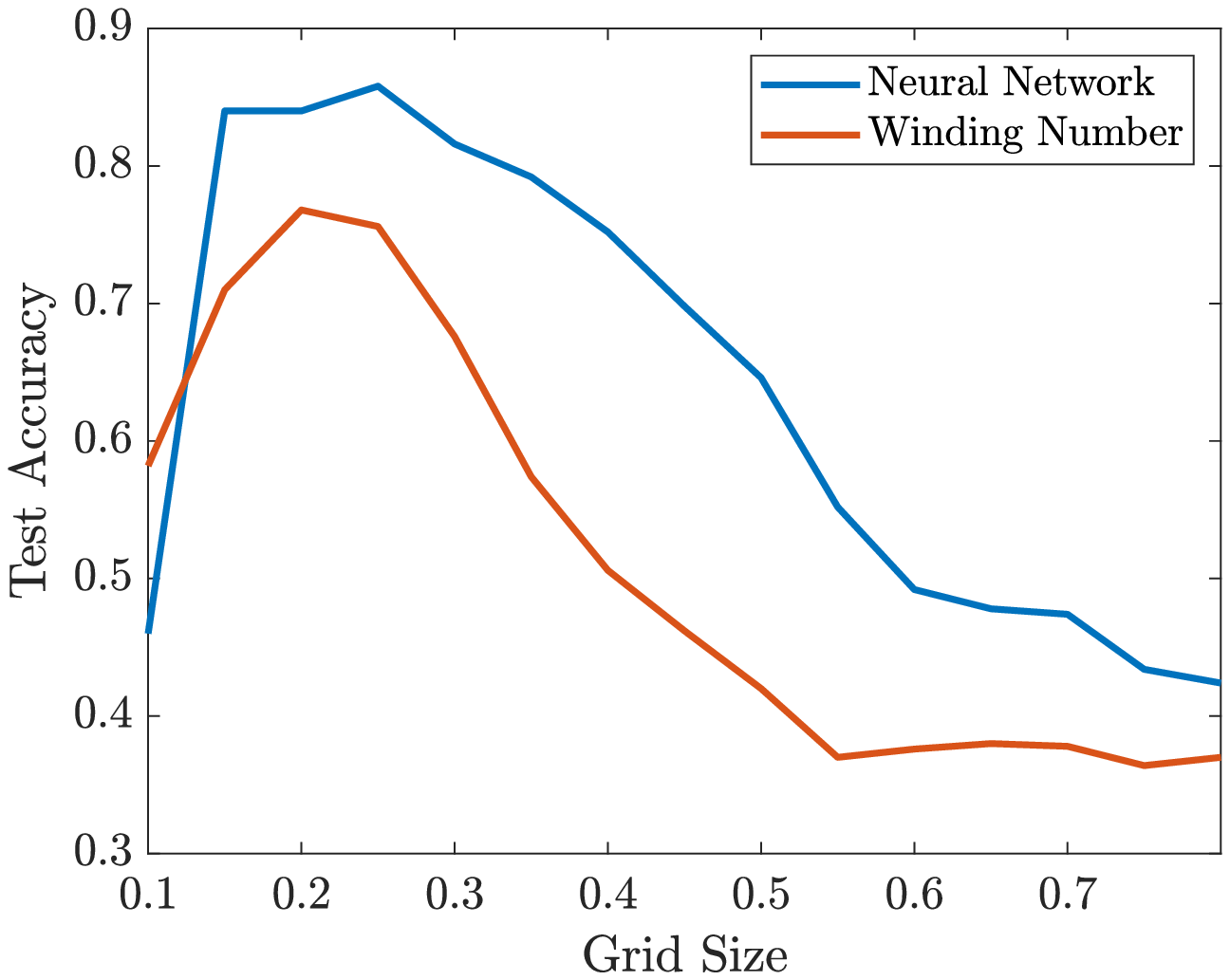}
    \end{center}
\caption{{\bf Machine learning model is less sensitive to system parameters than winding number.}
Classification accuracy vs. grid size for the CNN and the winding number method.}
\label{acc_v_grid}
\end{figure}

\bibliography{references}